\documentclass[letterpaper,twocolumn,prl,aps,showpacs,10pt,floatfix]{revtex4-1}

\usepackage[pdftex]{graphicx} 
\usepackage{epstopdf}
\usepackage{verbatim}
\usepackage{color}
\usepackage{subfigure}

\newcommand{\be}{\begin{equation}}
\newcommand{\ee}{\end{equation}}

\begin{document}
\title{Fractionalized charge excitations in a spin liquid on partially-filled pyrochlore lattices}
\author{Gang Chen$^1$, Hae-Young Kee$^{1,2}$, and Yong Baek Kim$^{1,2,3}$}
\affiliation{${}^1$Department of Physics, University of Toronto, Toronto, Ontario, M5S1A7, Canada}
\affiliation{${}^2$Canadian Institute for Advanced Research/Quantum Materials Program, Toronto, Ontario MSG 1Z8, Canada}
\affiliation{${}^3$School of Physics, Korea Institute for Advanced Study, Seoul 130-722, Korea}
\date{\today}

\begin{abstract}
We study the Mott transition from a metal to cluster Mott insulators in the 1/4- and 1/8-filled pyrochlore lattice systems. 
It is shown that such Mott transitions can arise due to charge localization in clusters or in tetrahedron units, 
driven by the nearest-neighbor repulsive interaction. The resulting cluster Mott insulator is a quantum spin liquid 
with a spinon Fermi surface, but at the same time a novel fractionalized charge liquid with charge excitations 
carrying half the electron charge. There exist two emergent U(1) gauge fields or ``photons" that mediate interactions 
between spinons and charge excitations, and between fractionalized charge excitations themselves, respectively. 
In particular, it is suggested that the emergent photons associated with the fractionalized charge excitations can 
be measured in X-ray scattering experiments. Various other experimental signatures of the exotic cluster 
Mott insulator are discussed in light of candidate materials with partially-filled bands on the pyrochlore lattice.
\end{abstract}

\date{\today}


\maketitle

In Mott insulators, strong correlation causes the charge localization\cite{Imada98}. As the charge excitation gap becomes
smaller near the insulator-metal transition, the strong local charge fluctuations can generate significant long-range
and/or ring exchange spin interactions. It has been recognized that these interactions may stabilize the so-called
quantum spin liquid (QSL)\cite{Motrunich05,Lee05}, where there exist charge-neutral spin-1/2 excitations or spinons
while spinless charge excitations are gapped\cite{Balents10}. 
In particular, when the transition from a metal to the spin liquid is continuous,
the resulting spin liquid may form a Fermi surface of the spinons. In the study of a Hubbard model at the $\frac{1}{2}$ 
filling for the 2D triangular and 3D hyperkagome lattices\cite{Senthil08July1,Senthil08July2,Podolsky09},
this new type of Mott transition is shown to occur as one increases the on-site Hubbard interaction.
On the experimental front, such transitions can be of relevance to QSL candidate materials such as the 2D triangular lattice
organic compound $\kappa$-(ET)$_2$Cu$_2$(CN)$_3$\cite{Shimizu03}
and the 3D hyperkagome material Na$_4$Ir$_3$O$_8$\cite{Okamoto07}.
In this spin liquid state, the spinons are interacting with an emergent U(1) gauge field
or ``photon", hence it is called the U(1) QSL\cite{Motrunich05,Lee05,Senthil08July1,Senthil08July2,Podolsky09,Chen13}.
On the other hand, the charge excitations behave trivially
and are simply localized on the lattice sites forming a charge ``solid" with gapped charge 
$q_e$ excitations. One may wonder whether it is possible to have a Mott insulator where the 
charge physics becomes non-trivial in addition to the spin sector. 

In this letter, we study Mott insulators and Mott transitions in partially-filled pyrochlore lattice systems. 
We uncover a novel cluster Mott insulator (CMI), where the electrons are localized within the tetrahedral clusters rather than on lattice sites. An example of CMI on the Kagome lattice 
has recently been discovered in LiZn$_2$Mo$_3$O$_8$ and studied by us theoretically\cite{Sheckelton12,*Sheckelton14,*Mourigal14,Chen14}. 
The ground state of the CMI on the pyrochlore lattice is shown to be a quantum spin liquid where 
there exist fractionalized charge excitations in addition to gapped spinons, and two kinds of emergent
gauge photons. Although the notion of charge fractionalization has been proposed in certain 
classically degenerate systems\cite{Fulde2002}, the charge fractionalization discussed in this paper is 
fundamentally different and is an intrinsic quantum effect. 
Besides the fundamental interest, this problem is of interest from the experimental point of view. 
Pyrochlore lattice systems with partially-filled bands occur in various materials with mixed-valence magnetic 
ions\cite{Urano00,Elmeguid04,Hirai13,Radaelli02}.
The model and underlying physics discussed in our work would potentially be 
relevant to such systems.

We focus on a single-band Hubbard model, 
\begin{eqnarray}
H &=& - t \sum_{\langle ij \rangle,\sigma} (c^{\dagger}_{i\sigma} c^{\phantom\dagger}_{j\sigma} + h.c.) 
-\mu \sum_i n_i \cr
 &&+ V\sum_{\langle ij \rangle} n_i n_j + \frac{U}{2} \sum_i (n_{i} - \frac{1}{2} )^2, 
\end{eqnarray}  
where 
$c^{\dagger}_{i\sigma}$ ($c^{\phantom\dagger}_{i\sigma}$) is the electron creation (annihilation)
operator at site $i$ with spin $\sigma$ and $n_{i}$ ($n_i = \sum_{\sigma }n_{i\sigma}$) 
is the electron number operator. We consider 
$\frac{1}{4}$- and $\frac{1}{8}$-filled cases.
Throughout this paper, we assume that the on-site Hubbard $U$ is the biggest 
energy scale. Notice, however, that 
 the interaction $U$ cannot cause electron localization for a
partially-filled band. It is the nearest-neighbor repulsion $V$ that  
drives the charge localization and the formation of Mott insulators.
Similarly to the $\frac{1}{2}$-filled case, the spin sector may form a QSL with a spinon Fermi surface for sufficiently large $V$.
In contrast to the $\frac{1}{2}$-filled case, however, the electrons in the Mott regime are localized on 
tetrahedral clusters with two(one) electrons per tetrahedron in the $\frac{1}{4}$($\frac{1}{8}$)-filled case.
In the classical $V=\infty$ limit, the electron site-occupation configurations of this CMI 
are highly degenerate\cite{Runge2004}. This is analogous to the degenerate ground-state manifold in the
classical spin ice\cite{Henley10} ($\frac{1}{2}$-magnetization plateau state\cite{Bergman06}) 
for the $\frac{1}{4}$ ($\frac{1}{8}$)-filled case.  

It is shown that, at finite $V$, the charge sector supports an additional emergent U(1) gauge field 
and fractionalization of charge quantum number in analogy to 
the quantum spin ice or $\frac{1}{2}$-magnetization plateau state. 
Therefore, the charge sector of the CMI is a U(1) fractionalized charge liquid (FCL). 
We show that the electron in this CMI fractionalizes into a
fermionic spinon and two charge bosons that carry {\it half the electron charge}.
The transition to a Fermi liquid metal occurs when the fractionally-charged bosons condense. 
We also discuss thermodynamic and spectrascopic properties 
of this novel Mott insulating phase.    

\begin{figure}[tp]
\subfigure[]{\label{fig1:a}\includegraphics[width=3.75cm]{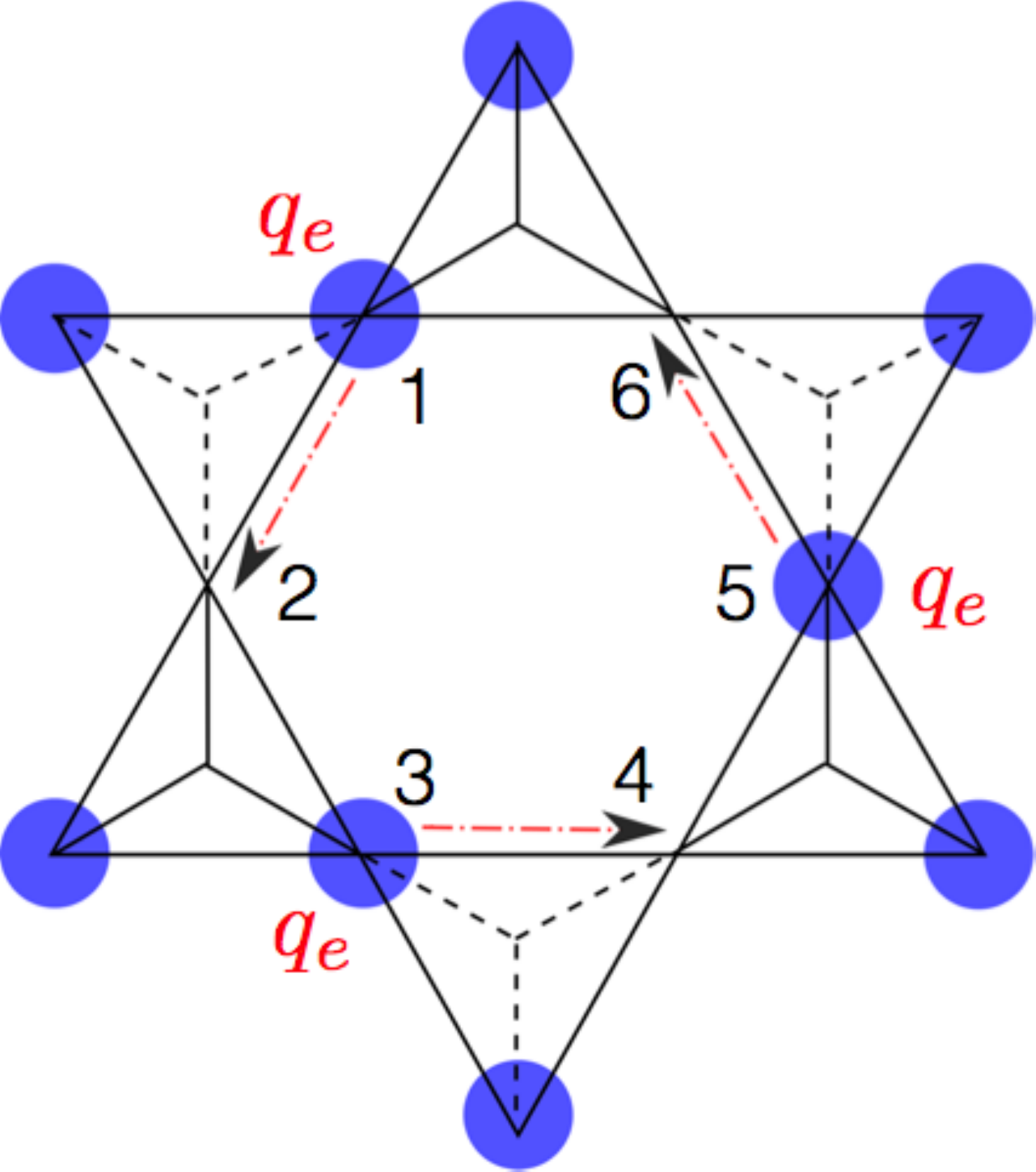}}
\hspace{0.8cm}
\subfigure[]{\label{fig1:b}\includegraphics[width=3.55cm]{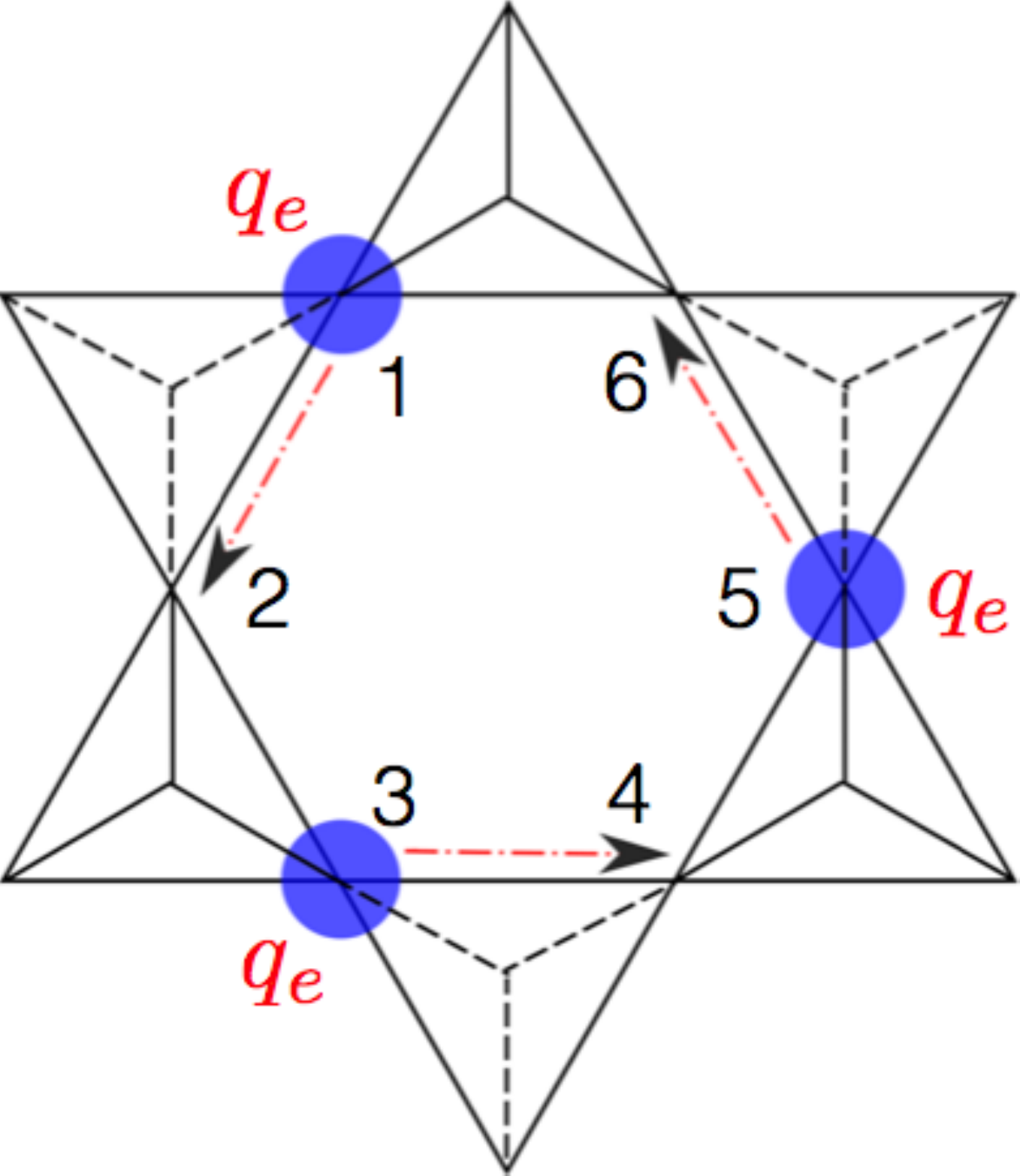}}
\subfigure[]{\label{fig1:c}\includegraphics[width=3.75cm]{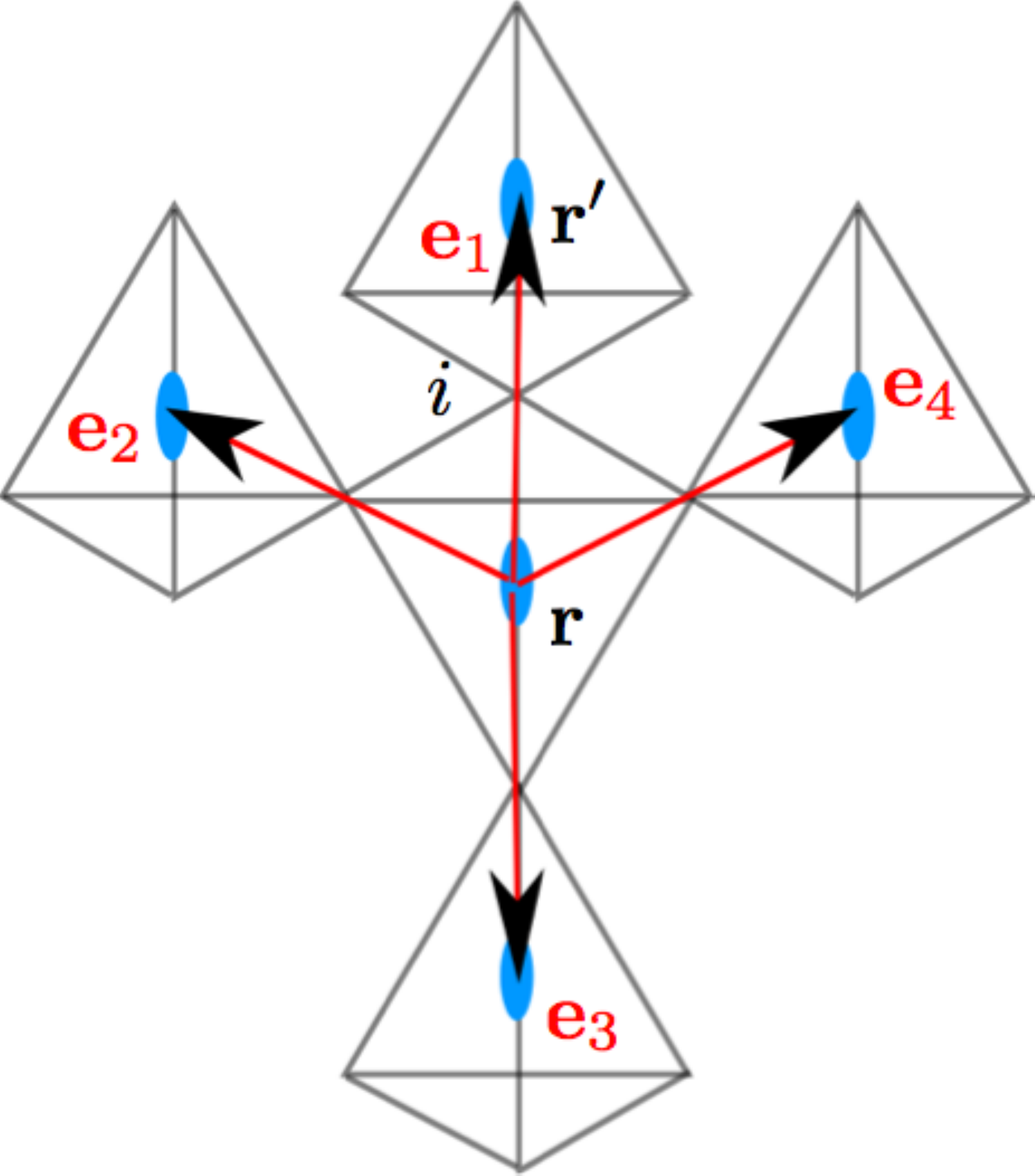}}
\hspace{0.8cm}
\subfigure[]{\label{fig1:d}\includegraphics[width=3.9cm]{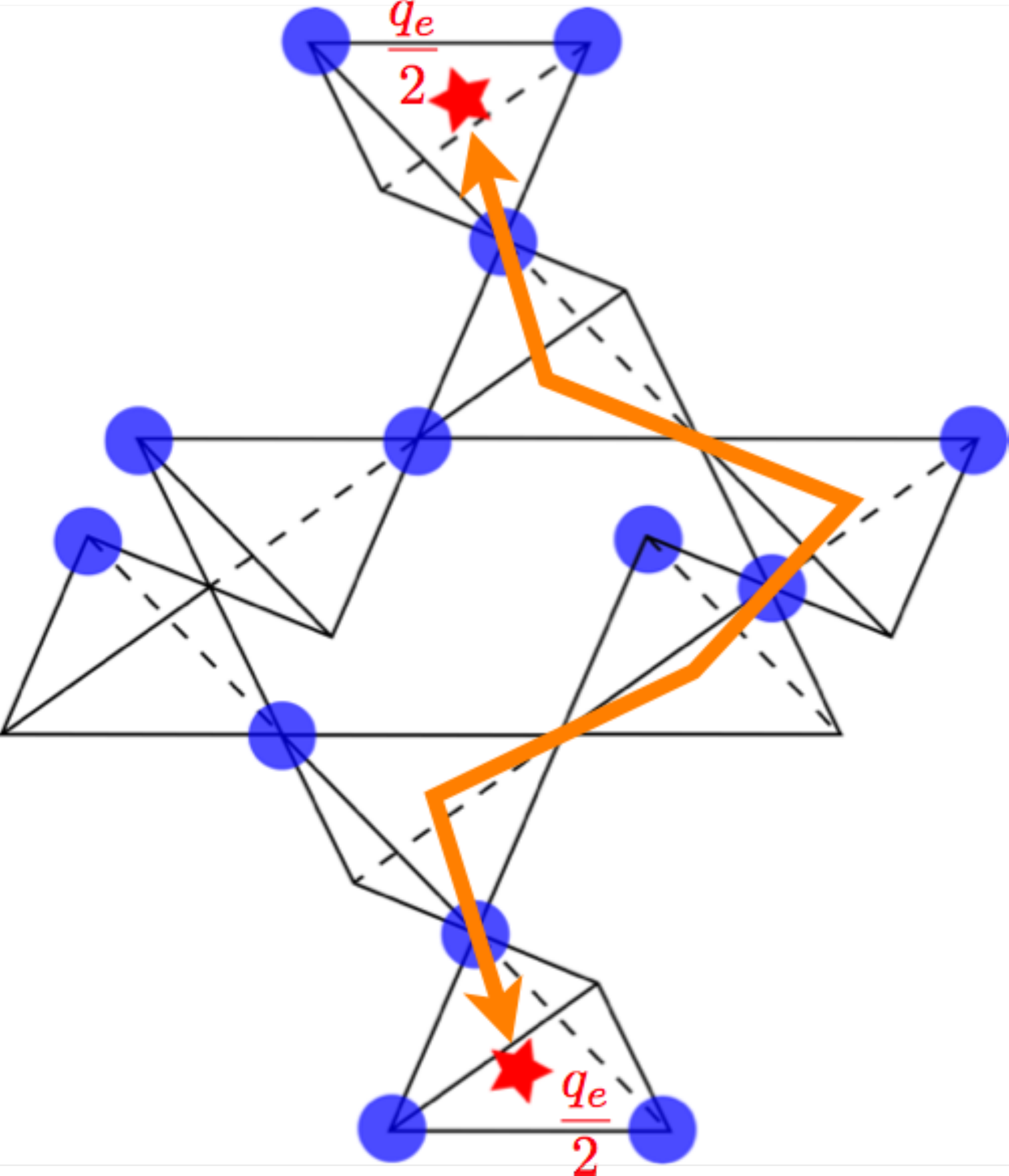}}
\subfigure[]{\label{fig:fig2}\includegraphics[width=8cm]{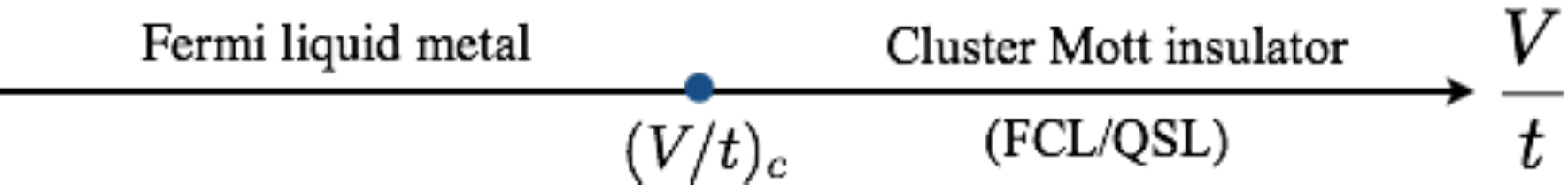}}
\caption{(Color online) The ring hopping processes of charge rotors around a hexagon 
in the CMI, for the $\frac{1}{4}$- and $\frac{1}{8}$-filled cases shown in (a) and (b). 
As shown in (c), ${\bf r}$ and ${\bf r}'$ are located on the center of the tetrahedra and 
form a dual diamond lattice. We use ``${\bf r},{\bf r}'$" (``$i,j$") to label 
the diamond (pyrochlore) lattice sites. In (c), ${\bf r} \in \text{A}$ diamond sublattice and 
${\bf e}_{\mu}$ are four vectors connecting A sublattice sites to the four neighboring B sublattice sites. 
In (d),  the electron charge fractionalization in the FCL/QSL phase is illustrated. The two end charge 
defects are connected by a fictitious string. 
The phase diagram at the $\frac{1}{4}$- or $\frac{1}{8}$-filling is plotted in (e). 
Here, $(\frac{V}{t})_c = 1.65$($0.98$) for the $\frac{1}{4}$($\frac{1}{8}$)-filling in the mean-field theory.
There are only two phases: a Fermi liquid metal and a CMI (FCL/QSL).}
\label{fig:fig1}
\end{figure}

\emph{Weak Mott regime for $\frac{1}{4}$ filling.} We start with the $\frac{1}{4}$ filled case.  
The model has a Fermi liquid ground state for $V \ll t$ \cite{flatness} 
and a Mott insulating ground state for $V \gg t$.  
To study the Mott transition of this Hubbard model, we first introduce the usual slave rotor
formalism\cite{Florens04,Lee05} and express the electron operator as
$c^{\dagger}_{i \sigma} =e^{i \theta_i} f^{\dagger}_{i\sigma}$, 
where $e^{i \theta_i}$ is the bosonic rotor operator carrying electric charge $q_e$ 
and $f^{\dagger}_{i\sigma}$ is the charge-neutral fermionic spinon operator.
To preserve the physical Hilbert space, we impose the gauge constraint 
$L_i^z = (\sum_{\sigma} f^{\dagger}_{i \sigma}
f^{\phantom\dagger}_{i\sigma} ) - \frac{1}{2}$, where 
$L_i^z$ is the conjugate operator of $\theta_j$ with 
$[\theta_i, L_j^z]= i \delta_{ij}$. 
Via a decoupling of the electron hopping term, the original Hubbard model is 
reduced to two coupled Hamiltonians $H_{\text{sp}}$ and $H_{\text{ch}}$ for the spin and 
charge sectors, respectively,
\begin{eqnarray}
H_{\text{sp}} &=& - \sum_{\langle ij \rangle, \sigma}  t^{\text{eff}}_{ij}  
(f^{\dagger}_{i\sigma} f^{\phantom\dagger}_{j\sigma} + h.c. )
- \sum_{i,\sigma}  (\mu + h_i)  f^{\dagger}_{i \sigma} f^{\phantom\dagger}_{i\sigma} 
\label{eq:Hs}
\\
H_{\text{ch}} &=& - \sum_{\langle ij \rangle} {J}^{\text{eff}}_{ij}  ( e^{i\theta_i -i \theta_j} + h.c.) 
+ V \sum_{\langle ij \rangle} L_i^z L_j^z  \nonumber 
\\ &&+ 3V \sum_i L_i^z + \sum_i h_i (L_i^z +\frac{1}{2}) + \frac{U}{2} \sum_i (L^z_i)^2 . 
\label{eq:Hc}
\end{eqnarray}
Here, $t^{\text{eff}}_{ij} = t \langle e^{i\theta_i - i\theta_j} \rangle \equiv | t^{\text{eff}}_{ij} | e^{ i a_{ij} } $, 
$ {J}^{\text{eff}}_{ij} = t \sum_{\sigma} \langle f^{\dagger}_{i\sigma} 
f^{\phantom\dagger}_{j\sigma}\rangle  \equiv  |  {J}^{\text{eff}}_{ij}  | e^{ - i a_{ij} }   $ and $h_i $ is the 
Lagrange multiplier that imposes the Hilbert space constraint.
With this reformulation of the Hubbard model, the Hamiltonians 
$H_{\text{sp}}$ and $H_{\text{ch}}$ are now invariant under an internal U(1) gauge transformation
$f_{i\sigma}^{\dagger} \rightarrow f_{i\sigma}^{\dagger} e^{-i \chi_i}, \theta_i \rightarrow
\theta_i + \chi_i$ and $a_{ij} \rightarrow a_{ij} + \chi_i - \chi_j$.
This internal U(1) gauge structure will be referred as U(1)$_{\text{sp}}$ in the following. 

In the half-filled case, the electrons are localized on the lattice sites in the Mott insulator. 
In the slave rotor formulation, the QSL Mott insulator corresponds 
to the deconfined phase of the U(1)$_{\text{sp}}$ gauge theory, and its transition to the metallic 
phase is induced by the condensation of the charge rotor\cite{Florens04,Lee05}. 
The situation for $\frac{1}{4}$ filling is somewhat different, 
even though the spin sector behaves similarly and forms a U(1)$_{\text{sp}}$ QSL with a 
spinon Fermi surface in the Mott regime. For the charge sector, the strong inter-site repulsion
$ \frac{V}{2} \sum_{\text{tet}} (\sum_{i \in \text{tet}} L_i^z )^2 + \text{const}$ (where tet refers to a tetrahedron) 
penalizes single charge motion from one tetrahedral cluster to another and leads to 
charge localization on the cluster. Hence, the total charge number on each tetrahedra 
is constrained to be two, or equivalently, satisfies the ``{\it charge ice constraint}" $\sum_{i \in \text{tet}} L_i^z = 0 $,
which is reminiscent of the spin ice constraint in the classical spin ice\cite{Bramwell01,Castelnovo08,Gingras09,Hermele04,Henley10,Savary12}. 
Similarly to the classical spin ice \cite{Henley10}, the 
classical charge ice configurations in the infinite $V$ limit are macroscopically degenerate\cite{Fulde2002,Runge2004}. 
These features drastically modify the charge sector physics. 

We now adopt a self-consistent mean-field approach and assume a uniform slave-rotor 
mean-field solution such that 
$t^{\text{eff}}_{ij}  \equiv t^{\text{eff}}$, ${J}^{\text{eff}}_{ij}  \equiv {J}^{\text{eff}}$ 
and $h_i \equiv h$. 
In the CMI, the rotor hopping 
$J^{\text{eff}}$ introduces quantum fluctuations and lifts the extensive charge ice degeneracy,
which is captured by a standard perturbative treatment of $J^{\text{eff}}$. 
We preserve the charge ice constraint in the ground state and obtain an effective ring rotor hopping model
from the third-order degenerate perturbation theory, 
\begin{eqnarray}
H_{\text{ch},\text{eff}} & = & - J_{\text{ring}} \sum_{\text{hexagon}} 
\cos (\theta_1-\theta_2 + \theta_3-\theta_4+\theta_5-\theta_6 ) 
\nonumber  \\
&&
+ \frac{U}{2} \sum_i ( L_i^z )^2,
\label{eq:eqHring}
\end{eqnarray}
where $J_{\text{ring}} = { 24 ({J}^{\text{eff}} )^3 }/{V^2}$ is 
the ring rotor-hopping amplitude around a hexagon plaquette (see Fig.~\ref{fig1:a}). 
This low-energy effective model acts on the charge ice manifold
and is analogous to the one obtained in the context 
of the quantum spin ice in the XXZ model\cite{Hermele04} on the pyrochlore lattice except 
that we have a large and finite interaction $U$ and $L^z$ can take the values 
of $\pm \frac{1}{2}$ and $\frac{3}{2}$ at the lattice length scale.  
Despite these small differences, the current model does share the same internal symmetries
as the quantum spin ice models and thus the universal properties of our model 
$H_{\text{ch},\text{eff}}$ is identical to the quantum spin ice in the low energy limit 
with $L^z =\pm \frac{1}{2}$. 
Therefore, the ground state of the charge sector is a U(1) {\it quantum charge ice}. 
The low energy U(1) gauge structure is obtained by
introducing lattice electric field $L_i^z \sim { E}_{{\bf r}{\bf r}'} $ and lattice vector 
potential $e^{i\theta_i} \sim e^{i {A}_{{\bf r}{\bf r}'}}$,
where ${\bf r}$(${\bf r}'$) lies on the A(B) diamond sublattice (Fig.~\ref{fig1:c})
and ${ E}_{{\bf r}{\bf r}'} = - { E}_{{\bf r}'{\bf r}},  
{A}_{{\bf r}{\bf r}'} = -{A}_{{\bf r}'{\bf r}}$.
To distinguish it from the U(1)$_{\text{sp}}$ gauge field, we label this 
as U(1)$_{\text{ch}}$ gauge field for the charge sector. 
The CMI is in the deconfined phase of this compact U(1)$_{\text{ch}}$ gauge theory
and we expect a gapless and linearly dispersing U(1)$_{\text{ch}}$ gauge photon to appear at low energies. 

Beyond the low energy regime, the rotor operator $e^{i\theta_i}$ creates a gapped charge-$q_e$ 
excitation that violates the charge ice constraints on the two neighboring tetrahedra centered 
at ${\bf r}$ and ${\bf r}'$. Just like the spin-$\frac{1}{2}$ bosonic spinon excitations in quantum spin ice,
this defect charge-$q_e$ excitation can be separated into two deconfined charge bosons ($\Phi^\dagger$) 
in arbitrary distances, each carrying half the electron charge.   
Therefore, the quantum charge ice state is a U(1) fractionalized charge liquid (FCL). 
As shown in Table.~\ref{tab:tab1}, these two fractionally 
charged bosons also carry the U(1)$_{\text{sp}}$ gauge charge ($Q^{\text{sp}}$) 
and U(1)$_{\text{ch}}$ gauge charge ($Q^{\text{ch}}$). 
Here the U(1)$_{\text{sp}}$ gauge charge is defined on the pyrochlore lattice site as 
$Q^{\text{sp}}_i = \sum_{\sigma} f^{\dagger}_{i \sigma} f^{\phantom\dagger}_{i\sigma} - L_i^z $
and the local U(1)$_{\text{ch}}$ gauge charge is defined on
the dual diamond lattice site (see Fig.~\ref{fig1:c}) as 
$Q^{\text{ch}}_{\bf r} = \eta_{\bf r} \sum_{\mu} L^z_{ {\bf r},{\bf r} + \eta_{\bf r} {\bf e}_{\mu} } $
where $\eta_{\bf r} = +1$($-1$) for ${\bf r}$ on the A(B) sublattice of the dual diamond lattice 
and ${\bf e}_{\mu}$ are the four nearest-neighbor vectors from the A sublattice sites (see Fig.~\ref{fig1:c}). 
The charge-$\frac{q_e}{2}$ bosons are fully gapped in the Mott insulator. 
As the electron hopping $t$ increases, the charge excitation gap becomes smaller and the 
charged bosons  condense upon closing the gap. 
The condensation of charge-$\frac{q_e}{2}$ bosons would make the two internal
gauge fields (U(1)$_{\text{sp}}$ and U(1)$_{\text{ch}}$) massive simultaneously, 
and drives a phase transition to a 
Fermi liquid metal (see Fig.~\ref{fig:fig2}).  
Therefore, there are only two phases in the phase diagram (see Fig.~\ref{fig:fig2}),
which is consistent with the quantum Monte Carlo simulation results for an interacting 
hardcore boson model at the half-filling on the pyrochlore lattice\cite{Isakov08}.

\begin{table}[htp]
\centering
\begin{tabular}{c|ccc} 
\hline
Operator & $Q^{\text{em}}$ & $Q^{\text{sp}}$  & $Q^{\text{ch}}$ 
\\
\hline
$c^{\dagger}_{i\sigma}$ & $q_e$ & 0 & 0 
\\
\hline
$f^{\dagger}_{i\sigma}$ & 0 & 1 & 0 
\\
\hline
$e^{i\theta_i}$ & $q_e$ & $-1$ & 0 
\\
\hline
$\Phi^{\dagger}_{\bf r}$, ${\bf r}\in$ A  & ${q_e}/{2}$ & $-{1}/{2}$ & 1
\\
\hline
$\Phi^{\dagger}_{\bf r}$, ${\bf r}\in$ B  & $- q_e/2$  &  $ 1/2$ & 1 
\\
\hline
\end{tabular}
\caption{Different kinds of gauge charges carried by various excitations. 
$Q^{\text{em}}$, $Q^{\text{sp}}$ and $Q^{\text{ch}}$ refer to the electric charge, 
U(1)$_{\text{sp}}$ gauge charge and U(1)$_{\text{ch}}$ gauge charge, respectively. 
$q_e$ is the charge of the electron.}
\label{tab:tab1}
\end{table}

In order to clearly represent both the U(1)$_{\text{ch}}$ gauge structure and charge fractionalization,
and to study the boson condensation transition for the charge sector $H_{\text{ch}}$, 
we employ the parton-gauge construction that was recently applied 
to the quantum spin ice\cite{Savary12,Savary13,Huang14,Sungbin12}. 
We include both the fractionalized charge bosons and a gauge field in the rotor variable as
$e^{i \theta_i} = \Phi^{\dagger}_{\bf r} \Phi^{\phantom\dagger}_{{\bf r}'} l_{{\bf r}{\bf r}'}^+, L^z_i = l^z_{{\bf r}{\bf r}'}$, 
where the pyrochlore lattice site $i = {\bf r}+ \frac{{\bf e}_\mu}{2}$ is the mid-point of the link $({\bf r}{\bf r}')$ on the dual diamond lattice
and ${\bf r}$(${\bf r}' = {\bf r} + {\bf e}_{\mu}$) belongs to the A(B) diamond sublattice. 
Here, $l^z_{{\bf r}{\bf r}'} \equiv {E}_{{\bf r}{\bf r}'}$ and $l_{{\bf r}{\bf r}'}^{\pm} 
\equiv \Delta_{{\bf r}{\bf r}'} e^{\pm i {A}_{{\bf r} {\bf r}' }}$ ($ \Delta_{{\bf r}{\bf r}'} \equiv | l^{\pm}_{{\bf r}{\bf r}'} |$) represent 
the lattice U(1)$_{\text{ch}}$ gauge fields on the links of the dual diamond lattice. 
To constrain the enlarged Hilbert space, we need $[\Phi_{\bf r}^{\phantom\dagger}, 
Q^{\text{ch}}_{\bf r} ] = \Phi_{\bf r}^{\phantom\dagger}$ and $[\Phi^{\dagger}_{\bf r}, 
Q^{\text{ch}}_{\bf r}] = - \Phi^{\dagger}_{\bf r}$. 
Now it is clear that the electron in the CMI fractionalizes into two charge-$\frac{q_e}{2}$ bosons and
a fermionic spinon (with an open string operator $l_{{\bf r} ,{\bf r}+ {\bf e}_{\mu}}^+$ connecting two bosons, see Fig.~\ref{fig1:d}),
\begin{equation}
c^{\dagger}_{ {\bf r}+ \frac{ {\bf e}_{\mu} }{2},  \sigma} = 
f^\dagger_{ {\bf r}+\frac{ {\bf e}_{\mu}}{2},\sigma}
\Phi_{\bf r}^{\dagger} \Phi_{  {\bf r}+ {\bf e}_{\mu} }^{\phantom\dagger}     
 l_{{\bf r}, {\bf r}+ {\bf e}_{\mu}}^+,
\end{equation}
where ${\bf r}\in$A sublattice. 
With the above construction, the charge sector Hamiltonian can be written as
\begin{eqnarray}
H_{\text{ch}} &=& - 
{J}^{\text{eff}} \sum_{{\bf r}, \mu \neq \nu} 
\Phi^{\dagger}_{{\bf r} + \eta_{\bf r} {\bf e}_{\mu} }  
\Phi^{\phantom\dagger}_{{\bf r}+\eta_{\bf r}
{\bf e}_{\nu}} 
l^{-\eta_{\bf r}}_{ {\bf r}, {\bf r}+\eta_{\bf r} {\bf e}_{\mu} } 
l^{ + \eta_{\bf r}}_{{\bf r},{\bf r}+\eta_{\bf r} {\bf e}_{\nu}} 
\nonumber \\
&&+ \frac{V}{2} \sum_{\bf r} (Q^{\text{ch}}_{\bf r})^2,
\end{eqnarray}
where we have dropped the linear $L^z$ term because of the emergent particle-hole 
symmetry at the Mott transition. The above charge Hamiltonian describes the minimal coupling 
of the fractionally charged bosons with the emergent U(1)$_{\text{ch}}$ gauge field on 
the dual diamond lattice. Within the gauge mean-field approximation\cite{Savary12}, 
we show that the Mott transition occurs at $(V/J^{\text{eff}})_c \approx 5.21$, where the charge bosons 
develop an energy gap. In this calculation, we have treated $L^z$ and $l^z$ as spin-$\frac{1}{2}$ variables,
which is a good approximation since double occupancy (or $L^z=\frac{3}{2}$) configuration is 
strongly suppressed by the large on-site interaction $U$. Together with the self-consistent 
mean-field theory for $H_{\text{sp}}$, we obtain a continuous Mott transition at 
$(V/t)_c \approx 1.65$ (see Fig.~\ref{fig:fig2}).

\emph{Weak Mott regime for $\frac{1}{8}$ filling.} For the CMI 
with $\frac{1}{8}$ eletron filling, the main difference is that the electron occupation number 
per tetrahedron is 1, {\it i.e.} $\sum_{i \in \text{tet} } L^z_i = - 1$. The low energy 
model of the charge sector is then obtained through the ring hopping processes of the rotors around 
a hexagon (see Fig.~\ref{fig1:b}). In the end, the charge occupation-number constraint and the low energy 
model are identical to the $\frac{1}{2}$-magnetization plateau state of a spin-$\frac{1}{2}$ XXZ model 
on the pyrochlore lattice in a uniform magnetic field\cite{Bergman06}.
It is known that the $\frac{1}{2}$-magnetization plateau state is a U(1) QSL with 
the same universal properties as the quantum spin ice\cite{Bergman06}. 
Therefore, the charge sector for the $\frac{1}{8}$-filled case is 
also a U(1)$_{\text{ch}}$ FCL with the same low energy excitations 
as the $\frac{1}{4}$-filled case.  

\emph{Strong Mott regime.} Here we turn to the strong Mott regime with $V\gg t$. 
Let us start with the CMI at the $\frac{1}{8}$-filling, where
the electrons on neighboring tetrahedra are always separated by one unoccupied site
(see Fig.~\ref{fig1:b}). The dominant interaction arises from the ring hopping processes of 
the three electrons on the hexagon and is described by 
\begin{eqnarray}
\label{eq:pert}
H_{\text{eff}} &=& - J_{\text{ring}}^{\text{e}} 
\sum_{\text{hexagon}}
\sum_{ \alpha \beta \gamma} 
( c^{\dagger}_{1\alpha} c^{\phantom\dagger}_{2\alpha}
c^{\dagger}_{3\beta} c^{\phantom\dagger}_{4\beta} 
c^{\dagger}_{5\gamma} c^{\phantom\dagger}_{6\gamma}
\nonumber \\
&& + c^{\dagger}_{1\alpha} c^{\phantom\dagger}_{6\alpha}
c^{\dagger}_{5\beta} c^{\phantom\dagger}_{4\beta} 
c^{\dagger}_{3\gamma} c^{\phantom\dagger}_{2\gamma}
+ h.c. ),
\end{eqnarray}
where $J_{\text{ring}}^{\text{e}}  = \frac{6t^3}{V^2}$ is the electron ring hopping amplitude. 
This interaction does not transfer charges between tetrahedra, but does transfer spin quantum numbers and 
hence overwhelms any other spin-spin interactions that arise from higher order processes. 
We emphasize that Eq.\ref{eq:pert} {\it cannot} be cast into the usual form of 
pairwise spin interactions or ring exchange, which is an important difference 
between the CMIs and conventional magnets. 
In conventional magnets, the spin moment can be considered as being coupled to  
a mean magnetic field generated by the exchange interactions from neighboring spins 
and if this mean magnetic field does not fluctuate strongly, the spin tends to 
align with this field and develop magnetic ordering. 
For the CMI here, such a mean magnetic field cannot be 
defined from the interaction in Eq.\ref{eq:pert} and thus we do not expect simple magnetic 
ordering. Then, for the spin sector, we may expect the QSL from the weak 
Mott regime to remain in the strong Mott regime. 
For the charge sector, we note that the effect of Eq.\ref{eq:pert} on the charge excitations
is identical to the charge rotor hopping processes in Eq.\ref{eq:eqHring}. 
Following the same reasoning as presented for the weak Mott regime, 
we expect the same U(1)$_{\text{ch}}$ FCL  to arise in the strong Mott regime. 
In other words, the quasi-itinerancy nature of the electrons inside the FCL helps the spin quantum numbers 
to freely propagate, which prevents simple magnetic ordering and may stabilize a QSL state. This quasi-itinerancy 
would be a new mechanism to stabilize quantum spin liquid phases, in addition to the known
mechanisms such as geometric frustration, low-dimensionality, and the proximity to Mott transitions.

In the strong Mott regime for the $\frac{1}{4}$-filling, there exists a  
superexchange spin-spin interaction between nearest neighbor sites with 
the exchange coupling $J_{\text{ex}} = \frac{4 t^2}{U-V} + \frac{8t^3}{V^2}$. 
Since this energy scale $J_{\text{ex}}$ is larger than or comparable to the electron ring 
hopping amplitude $J_{\text{ring}}^{\text{e}}$, the FCL/QSL may survive or be 
destabilized depending on different parameter regimes\cite{unpublished}.

\emph{Discussion.} We now discuss the experimental signatures related to these exotic 
CMIs. We begin with the principal physical properties in the vincinity of the Mott transition.
The Mott transition is continuous in the mean-field theory, but might turn to a weakly
first order transition upon including U(1)$_{\text{ch}}$ gauge fluctuations\cite{Halperin74}. 
Even in that case, the first order effect may be important only at extremely low temperatures. 
So for a rather wide temperature range, 
the physics near the Mott transition is controlled by the {\em critical} fractionalized charge 
bosons coupled to the U(1)$_{\text{ch}}$ and U(1)$_{\text{sp}}$ gauge fields, 
and the fermionic spinons coupled to the U(1)$_{\text{sp}}$ gauge field. 
Similarly to the half-filled case studied earlier\cite{Podolsky09}, the dynamical critical exponent
for the charge boson (fermionic spinon with U(1)$_{\text{sp}}$) is $z=1$ ($z=3$). 
Hence we expect two crossover temperature scales for specific heat and electric
resistivity, respectively. Due to further fractionalization 
of charge excitations, the tunneling density of states at the transition would be 
highly suppressed as $N_{\text{tunn}}^{\text{crit}}(\omega) \sim \omega^4$
instead of $\omega^2$ as in the half-filled case\cite{Podolsky09}.

The low energy U(1)$_{\text{ch}}$ gauge field originates from the 
electron charge fluctuations and may be probed by 
elastic and/or inelastic X-ray scattering. Similarly to the spin structure factor in the quantum spin ice\cite{Hermele04,Isakov08,Shannon2012,*Sikora09,*Benton2012,Savary12}, 
the inelastic charge structure factor of the CMI at low energies 
can be regarded as the emergent ``electric-field" correlator and is given by
$ \text{Im} [ {E}^\alpha_{-{\bf k},-\omega} {E}^{\beta}_{ {\bf k},\omega }  ] 
\propto [ \delta_{\alpha\beta}  - \frac{k_\alpha k_\beta}{{\bf k}^2} ] \, \omega \, \delta(\omega - v |{\bf k} |)$,
where ${{\bf E}}_{ {\bf r} + \frac{1}{2} {\bf e}_{\mu}} \equiv L^z_{ {\bf r}, {\bf r}+ {\bf e}_\mu } 
\frac{\bf{e}_\mu}{| {\bf e}_{\mu} |} = 
(n_{{\bf r} + \frac{1}{2} {\bf e}_{\mu}}  - \frac{1}{2} )\frac{\bf{e}_\mu}{| {\bf e}_{\mu} |} $
and ${\bf r} \in$ A diamond sublattice. Here $v$ is the speed of the U(1)$_{\text{ch}}$ gauge photon. 

The CMI is expected to lose the quantum coherence around a temperature 
$T^{\ast} \sim max[J^{\text{e}}_{\text{ring}}, J^{\text{ex}} ]$ in the Mott regime. 
In the temperature range $T^{\ast} \lesssim T \lesssim V$, the cluster electron occupation-number 
constraint still holds and the system is described by 
a thermal charge liquid, where degenerate charge configurations are equally allowed. 
Similarly to the classical spin ice\cite{Henley10}, the equal-time charge structure factor is given by
$\langle {E}^\alpha_{-{\bf k}} {E}^{\beta}_{ {\bf k} } \rangle \propto
\delta_{\alpha\beta}  - \frac{k_\alpha k_\beta}{{\bf k}^2} $, which leads to 
the pinch point structures in the ${\bf k}$ space \cite{Bramwell01,Castelnovo08,Gingras09,Henley10}. 

There exist several candidate materials for $\frac{1}{4}$- or $\frac{1}{8}$-filled pyrochlore lattice systems. 
Various spinels such as LiV$_2$O$_4$ (with V$^{3.5+}$:$d^{1.5}$)\cite{Urano00}, CuIr$_2$S$_4$ 
(with Ir$^{3.5+}$:d$^{5.5}$)\cite{Radaelli02} and  
GaTa$_4$Se$_8$ (with Ta$^{3.25+}$:$d^{1.75}$)\cite{Elmeguid04} may be good candidates 
for $\frac{1}{4}$- and $\frac{1}{8}$-filling cases. The $\beta$-pyrochlore system CsW$_2$O$_6$ (with W$^{5.5+}$: $d^{0.5}$)\cite{Hirai13} may also be a promising system where the physics discussed here can be explored.

\emph{Acknowledgements.} This work was supported by the NSERC, CIFAR, and 
Centre for Quantum Materials at the University of Toronto.
GC thanks Q. Si for the hospitality during his visit to Rice University 
when a related idea was motivated. YBK thanks S. Isakov for an earlier collaboration in related ideas.  
We also thank J.-H. Jiang, A. Burkov and 
A. Paramekanti for illuminating discussions.

\bibliography{refs}

\begin{thebibliography}{38}
\bibitem [{fla()}]{flatness}%
  \BibitemOpen
  \href@noop {} {}\bibinfo {note} {\textnormal{For the uniform nearest-neighbor
  hopping model, the Fermi level lies at a flat band extending from X to W in
  the Brillouin zone for the 1/4-filling case. In principle, this flat Fermi
  surface may cause other instabilities. For a generic tight-binding model (for
  example, with a spin-dependent hopping amplitude), such a flat band and
  instabilities may be avoided.}}\BibitemShut {Stop}%
\end{thebibliography}

\end{document}